\documentclass[epj]{webofc}
\usepackage[utf8]{inputenc}
\usepackage[varg]{txfonts}   
\usepackage{booktabs}
\usepackage{xcolor}
\definecolor{darkred}{rgb}{0.4,0.0,0.0}
\definecolor{darkgreen}{rgb}{0.0,0.4,0.0}
\definecolor{darkblue}{rgb}{0.0,0.0,0.4}
\usepackage[bookmarks,linktocpage,colorlinks,
    linkcolor = darkred,
    urlcolor  = darkblue,
    citecolor = darkgreen]{hyperref}
\usepackage{bm}
\wocname{EPJ Web of Conferences}
\woctitle{Lattice2017}
\newcommand{\para}{\parallel}
\newcommand{\leftI}{\left\lvert}
\newcommand{\midI}{\middle\vert}
\newcommand{\rightI}{\right\rvert}
\numberwithin{equation}{section}
\begin{document}
%
\selectlanguage{english}
\title{%
\boldmath
Semileptonic $B$-meson decays to light pseudoscalar mesons on the HISQ ensembles
}
\author{%
\firstname{Zechariah} \lastname{Gelzer}\inst{1,2}\fnsep\thanks{\email{zgelzer@illinois.edu}} \and
\firstname{C.} \lastname{Bernard}\inst{3} \and
\firstname{C.} \lastname{DeTar}\inst{4} \and
\firstname{A.X.} \lastname{El-Khadra}\inst{1,5} \and
\firstname{E.} \lastname{G\'amiz}\inst{6} \and
\firstname{Steven} \lastname{Gottlieb}\inst{7} \and
\firstname{Andreas S.} \lastname{Kronfeld}\inst{5,8}\fnsep\thanks{Speaker, \email{ask@fnal.gov}} \and
\firstname{Yuzhi} \lastname{Liu}\inst{7} \and
\firstname{Y.} \lastname{Meurice}\inst{2} \and
\firstname{J.N.} \lastname{Simone}\inst{5} \and
\firstname{D.} \lastname{Toussaint}\inst{9} \and
\firstname{R.S.} \lastname{Van de Water}\inst{5} \and
\firstname{R.} \lastname{Zhou}\inst{5}
}
\institute{%
Department of Physics,
University of Illinois,
Urbana, IL 61801, USA
\and
Department of Physics and Astronomy,
University of Iowa,
Iowa City, IA 52242, USA
\and
Department of Physics,
Washington University,
St. Louis, Missouri, 63130, USA
\and
Department of Physics and Astronomy,
University of Utah,
Salt Lake City, Utah 84112, USA
\and
Fermi National Accelerator Laboratory,
Batavia, Illinois, USA
\and
CAFPE and Departamento de F\'isica Te\'orica y del Cosmos,
Universidad de Granada,
Granada, Spain
\and
Department of Physics,
Indiana University,
Bloomington, Indiana 47405, USA
\and
Institute for Advanced Study,
Technische Universit\"at M\"unchen,
D-85748 Garching, Germany
\and
Department of Physics,
University of Arizona, Tucson,
Arizona 85721, USA
}
\abstract{%
We report the status of an ongoing lattice-QCD calculation of form factors
for exclusive semileptonic decays of $B$~mesons with both charged currents
($B\to\pi\ell\nu$, $B_s\to K\ell\nu$) and neutral currents
($B\to\pi\ell^+\ell^-$, $B\to K\ell^+\ell^-$). The results are important for
constraining or revealing physics beyond the Standard Model. This work uses
MILC's (2+1+1)-flavor ensembles with the HISQ action for the sea and light
valence quarks and the clover action in the Fermilab interpretation for the
$b$~quark. Simulations are carried out at three lattice spacings down to
$0.088$~fm, with both physical and unphysical sea-quark masses. We present
preliminary results for correlation-function fits.
}
\maketitle
\section{Introduction}\label{sec:intro}

\begin{figure}
    \centering
    \includegraphics[width=0.49\textwidth]{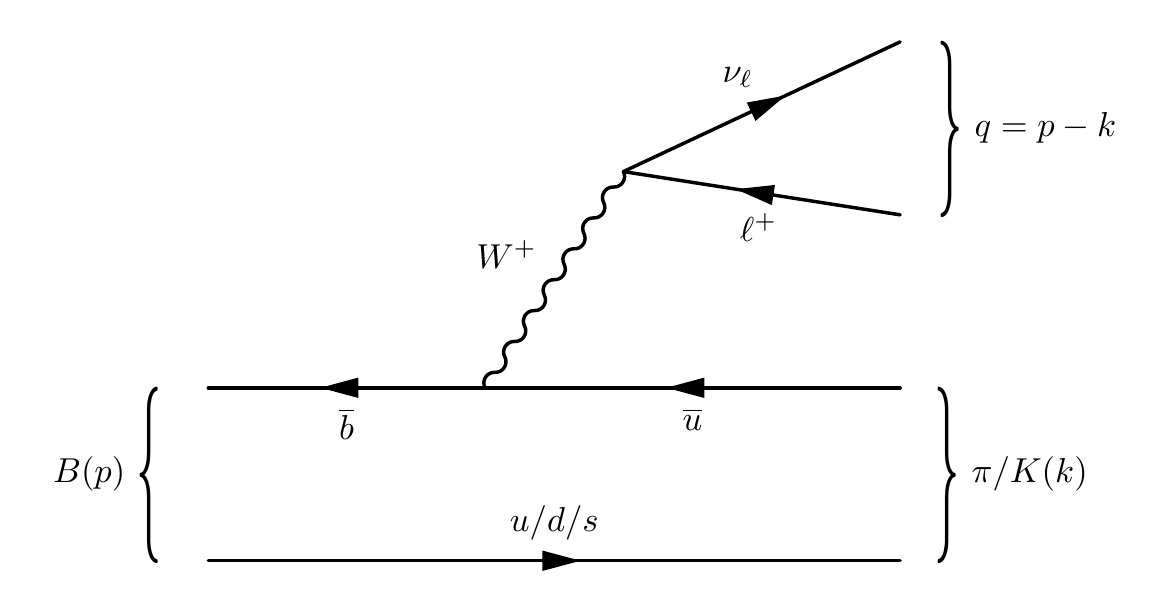}
    \hfill
    \includegraphics[width=0.49\textwidth]{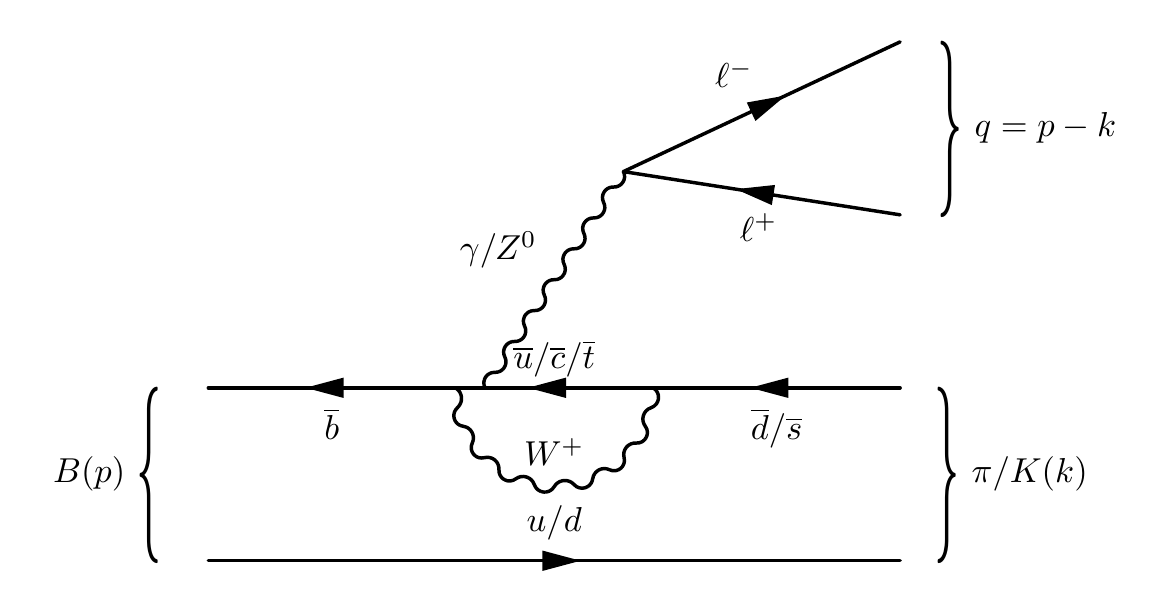}
    \caption{%
        Feynman diagrams for tree-level charged-current decays
        $B_{(s)}\to\pi(K)\ell\nu_\ell$ (left) and
        rare flavor-changing neutral-current interactions
        $B\to\pi(K)\ell^+\ell^-$ (right), with all QCD corrections suppressed.
    }
    \label{fig:feyn}
\end{figure}

Semileptonic decays of $B_{(s)}$~mesons are important to searches for new
physics at the intensity frontier.
Decays mediated at tree-level by charged currents (see Figure~\ref{fig:feyn},
left), namely $B\to\pi\ell\nu_\ell$ and $B_s\to K\ell\nu_\ell$, provide
precise determinations of the CKM matrix element $|V_{ub}|$. This element is
crucial for testing CKM unitarity and thus important in the search for new
physics. There are well-known discrepancies between inclusive and exclusive
determinations of $|V_{ub}|$ \cite{Patrignani:2016xqp}. Exclusive
determinations combine theoretical calculations of nonperturbative form
factors with experimental measurements of decay rates. This work seeks to
further reduce the uncertainties from theory.

Decays mediated by neutral currents (see Figure~\ref{fig:feyn}, right),
namely $B\to\pi\ell^+\ell^-$ and $B\to K\ell^+\ell^-$, provide precise tests
of the Standard Model. Since flavor-changing neutral-current (FCNC)
interactions are rare, they are sensitive to contributions from new physics.
Tension currently exists between the Standard Model and experimental data
pertaining to these FCNC interactions \cite{Du:2015tda}. The semileptonic
decays and neutral mixing of $B$~mesons have been studied in experiments at
\textsc{BaBar} \cite{Aubert:2005kf}, Belle \cite{Abe:2004mz}, CDF
\cite{Abulencia:2006ze}, D\O\ \cite{Abazov:2012zz}, and LHCb
\cite{Aaij:2013gja, Aaij:2014pli, Aaij:2015nea} and will become available
from ongoing measurements at LHCb and the upcoming Belle II experiment.
Errors from theory and experiment are commensurate and expected to be
improved.

In this work, we calculate form factors from first principles in lattice
QCD for the exclusive decays $B\to\pi\ell\nu_\ell$, $B_s\to K\ell\nu_\ell$,
$B\to\pi\ell^+\ell^-$, and $B\to K\ell^+\ell^-$. We use (2+1+1)-flavor MILC
HISQ ensembles, where the light quarks are at physical masses. The purpose of
this work is to improve theoretical confrontations with experiment, both by
determining $|V_{ub}|$ in charged-current decays and by comparing branching
fractions in neutral-current decays.

\section{Matrix elements and form factors}\label{sec:defs}

In the Standard Model and beyond, transitions between pseudoscalar mesons can
be mediated by a vector current $\mathcal{V}^\mu = \overline{q} \gamma^\mu b$,
a tensor current $\mathcal{T}^{\mu \nu} = i \overline{q} \sigma^{\mu \nu} b$,
or a scalar current $\mathcal{S} = \overline{q} b$. The resulting matrix
elements are as follows:
\begin{align}
    \left< P(k) \midI \mathcal{V}^\mu \midI B(p) \right>
        &= f_+(q^2) \left( p^\mu + k^\mu - \frac{M_B^2 - M_P^2}{q^2} q^\mu
        \right) + f_0(q^2) \frac{M_B^2 - M_P^2}{q^2} q^\mu ,
    \label{eq:matV} \\
        &= \sqrt{2 M_B} \left[ v^\mu f_\para(E_P) + k_\perp^\mu f_\perp(E_P)
        \right] ,
    \label{eq:matVproj} \\
    \left< P(k) \midI \mathcal{T^{\mu \nu}} \midI B(p) \right> &= f_T(q^2)
        \frac{2}{M_B + M_P} \left( p^\mu k^\nu - p^\nu k^\mu \right) ,
    \label{eq:matT} \\
    \left< P(k) \midI \mathcal{S} \midI B(p) \right> &= f_0(q^2)
        \frac{M_B^2 - M_P^2}{m_b - m_q} ,
    \label{eq:matS}
\end{align}
where $P=\pi,K$ is the light-pseudoscalar meson in the final state
and $k$ its four-momentum, $B=B,B_s$ is the $B_{(s)}$ meson in the
initial state and $p$ its four-momentum, and $q=p-k$ is the four-momentum
carried off by the leptons. In Eq.~(\ref{eq:matVproj}), $f_+$ and $f_0$ are
rewritten in terms of $f_\perp$ and $f_\para$, where $v=p/M_B$ is the
four-velocity of the $B$~meson, $k_\perp = k - (p \cdot v)\,v$ is the
projection of the four-momentum of the final-state meson in the direction
perpendicular to $v$, and $E_P = k\cdot v$ is the energy of the final-state
meson in the $B$-meson rest frame. Thus we may reconstruct $f_+(q^2)$ and
$f_0(q^2)$ as linear combinations of $f_\perp(E_P)$ and $f_\para(E_P)$:

\begin{align}
    f_+(q^2) &= \frac{1}{\sqrt{2 M_B}} \left[ f_\para(E_P) + \left( M_B - E_P
        \right) f_\perp(E_P) \right] ,
    \label{eq:fV} \\
    f_0(q^2) &= \frac{\sqrt{2 M_B}}{M_B^2 - M_P^2} \left[ \left( M_B - E_P
        \right) f_\para(E_P) + \left( E_P^2 - M_P^2 \right) f_\perp(E_P)
        \right] .
    \label{eq:fS}
\end{align}
Eqs.~(\ref{eq:fV}) and (\ref{eq:fS}) automatically satisfy the kinematic
constraint $f_+(0)=f_0(0)$. It becomes straightforward to obtain the form
factors $f_\perp$, $f_\para$, and $f_T$ as functions of the pion or kaon
energy by analyzing two- and three-point correlation functions in the
lattice-QCD calculation.

\section{Lattice-QCD calculation}\label{sec:lQCD}

We use (2+1+1)-flavor ensembles generated by the MILC Collaboration, which
include the dynamical effects of $u$, $d$, $s$, and $c$ quarks in the sea.
The isospin limit $m_u=m_d=m_l$ is used for the light quarks and the same
input masses are used for corresponding sea and valence quarks. There are
five ensembles at three lattice spacings down to $0.088$~fm and three of
these ensembles use physical quark masses. The highly-improved staggered
quark (HISQ) action is used for the sea and light valence quarks, while the
clover action in the Fermilab interpretation is used for the $b$ quark.
Gluons are implemented with the one-loop improved L\"uscher--Weisz action. To
determine the lattice scale, we use the gradient-flow quantity $w_0/a$
\cite{Borsanyi:2012zs}, whose continuum value is given by $w_0=0.1714(15)$~fm
\cite{Bazavov:2015yea}. Ensemble properties, including both parameter inputs
and elementary outputs, are given in Table~\ref{tab:ens}.

\begin{table}
    \small
    \centering
    \caption{%
        Parameter inputs and elementary outputs used in this work.
        Columns are (from left to right):
            approximate lattice spacing,
            lattice size,
            quark masses in lattice units,
            $b$-quark hopping parameter,
            number of configurations and different source times,
            product of pion mass with linear spatial dimension, and
            gradient-flow quantity.
    }
    \label{tab:ens}
    \begin{tabular}{*{10}{c}}
    \toprule
        $\approx a$~(fm) &
        $N_s^3 \times N_t$ &
        $a m_l^\prime$ &
        $a m_s^\prime$ &
        $a m_c^\prime$ &
        $\kappa_b^\prime$ &
        $N_\text{cfg} \times N_\text{src}$ &
        $T$ &
        $a M_\pi N_s$ &
        $w_0 / a$ \\
    \midrule
        $0.15$ &
        $32^3 \times 48$ &
        $0.00235$ &
        $0.0647$ &
        $0.831$ &
        $0.07732$ &
        $3630 \times 8$ &
        $14$ &
        $3.3$ &
        $1.1468(4)$ \\
    \midrule
        $0.12$ &
        $24^3 \times 64$ &
        $0.0102$ &
        $0.0509$ &
        $0.635$ &
        $0.08574$ &
        $1053 \times 8$ &
        $17$ &
        $4.5$ &
        $1.3835(10)$ \\
        $0.12$ &
        $32^3 \times 64$ &
        $0.00507$ &
        $0.0507$ &
        $0.628$ &
        $0.08574$ &
        $1000 \times 8$ &
        $17$ &
        $4.3$ &
        $1.4047(9)$ \\
        $0.12$ &
        $48^3 \times 64$ &
        $0.00184$ &
        $0.0507$ &
        $0.628$ &
        $0.08574$ &
        $986 \times 8$ &
        $17$ &
        $4.0$ &
        $1.4168(10)$ \\
    \midrule
        $0.088$ &
        $64^3 \times 96$ &
        $0.0012$ &
        $0.0363$ &
        $0.432$ &
        $0.09569$ &
        $925 \times 8$ &
        $25$ &
        $3.7$ &
        $1.9470(13)$ \\
    \bottomrule
    \end{tabular}
\end{table}

The lattice-QCD calculations in this work employ two- and three-point
correlation functions defined as follows:
\begin{align}
    C_2^B(t;\bm{0}) &= \sum_{\bm{x}} \left< \mathcal{O}_B(t,
        \bm{x}) \, \mathcal{O}^\dag_B(0, \bm{0}) \right> ,
    \label{eq:C2B} \\
    C_2^P(t;\bm{k}) &= \sum_{\bm{x}} \left< \mathcal{O}_P(t, \bm{x}) \,
        \mathcal{O}^\dag_P(0, \bm{0}) \right> e^{-i \bm{k} \cdot \bm{x}} ,
    \label{eq:C2P} \\
    C_3^{\mu(\nu)}(t, T;\bm{k}) &= \sum_{\bm{x}, \bm{y}} e^{i \bm{k}
        \cdot \bm{y}} \left< \mathcal{O}_P(0, \bm{0}) \, J^{\mu (\nu)}(t,
        \bm{y}) \, \mathcal{O}^\dag_B(T, \bm{x}) \right> ,
    \label{eq:C3}
\end{align}
where $J^{\mu(\nu)} = V^\mu, T^{\mu\nu}$ are the lattice currents. The
final-state-meson momenta are generated up to $\bm{k}=(4,\,0,\,0)\times
2\pi/(a N_s)$. Correlation functions are computed from multiple source
locations on each configuration to increase statistics. This is accomplished
by setting eight temporal source locations at $t=0,\,N_t/8,\,\dots,\,7N_t/8$,
while fixing the spatial source locations at $\bm{x} = \bm{0}$. The
autocorrelations between successive configurations are mitigated by binning,
wherein we analyze both the exponential autocorrelation time and the scaling
of variances with bin size. Three-point correlation functions are computed at
two source--sink separations $T$ and $T+1$ (see Table~\ref{tab:ens}).

Operators for currents on the lattice are related to those in the continuum
according to $\mathcal{J} \doteq Z_J J$. We use a mostly nonperturbative
matching,
\begin{align}
    Z_J = \rho_J \sqrt{Z_{V^4_{bb}} Z_{V^4_{qq}}} ,
    \label{eq:ZJ}
\end{align}
where $q=l,s$ for $P=\pi,K$, both square-root factors are computed
nonperturbatively, and $\rho_J$ are computed perturbatively. The
one-loop-corrected $\rho_J$ are not yet available in the literature. They
will be used to introduce a blinding procedure to our analysis, as in Refs.~\cite{Lattice:2015tia,Bailey:2015dka}.

\section{Analysis}\label{sec:analysis}

We extract the masses and energies of mesons on the lattice by fitting the
two-point correlation functions to these forms:
\begin{align}
    C_2^B(t;\bm{0}) &= \sum_{n=0}^{2N-1} (-1)^{n (t + 1)}
        \frac{\leftI Z_B^{(n)}\rightI^2}{2 M_B^{(n)}}
        \left[  e^{-M_B^{(n)} t} + e^{-M_B^{(n)} (N_t - t)} \right] ,
    \label{eq:C2fitB} \\
    C_2^P(t;\bm{k}) &= \sum_{n=0}^{2N-1} (-1)^{n (t + 1)}
        \frac{\leftI Z_P^{(n)}\rightI^2}{2 E_P^{(n)}}
        \left[  e^{-E_P^{(n)} t} + e^{-E_P^{(n)} (N_t - t)} \right] ,
    \label{eq:C2fitP}
\end{align}
where the overlap amplitudes are
$Z_B^{(n)}=\left< 0\midI\mathcal{O}_B\midI B^{(n)}\right>$ and
$Z_P^{(n)}=\left< 0\midI\mathcal{O}_P\midI P^{(n)}\right>$. We use $N=3$ in
all cases except for the zero-momentum pion, which does not contain states
with opposite parity. We loosely constrain the priors for amplitudes and
energies according to the procedures outlined in Ref.~\cite{Bazavov:2016nty}.
For the fit ranges $[t_\text{min}, t_\text{max}]$, which are listed in
Table~\ref{tab:tfit}, we choose $t_\text{min}$ in physical units to be
consistent for corresponding mesons on different ensembles, as well as to
maximize goodness of fit. We choose $t_\text{max}$ by requiring that the
fractional errors in the two-point correlators remain below $5\%$. Fit
results are confirmed to be stable under variations in $t_\text{min}$,
$t_\text{max}$, $N$, and prior widths.

\begin{table}
    \small
    \centering
    \caption{%
        Fit ranges $[t_\text{min}, t_\text{max}]$ for each meson in fits to
        two-point correlators, where $t_\text{max}$ is quoted at $\bm{k}=\bm{0}$.
    }
    \label{tab:tfit}
    \begin{tabular}{*{6}{c}}
    \toprule
        $\approx a$~(fm) &
        $a m_l^\prime ~/~ a m_s^\prime$ &
        $B$ &
        $B_s$ &
        $K$ &
        $\pi$ \\
    \midrule
        $0.15$ &
        $0.00235 ~/~ 0.0647$ &
        $[2,\, 15]$ &
        $[2,\, 24]$ &
        $[4,\, 23]$ &
        $[4,\, 23]$ \\
    \midrule
        $0.12$ &
        $0.0102 ~/~ 0.0509$ &
        $[3,\, 18]$ &
        $[3,\, 24]$ &
        $[5,\, 31]$ &
        $[5,\, 31]$ \\
        $0.12$ &
        $0.00507 ~/~ 0.0507$ &
        $[3,\, 17]$ &
        $[3,\, 26]$ &
        $[5,\, 31]$ &
        $[5,\, 31]$ \\
        $0.12$ &
        $0.00184 ~/~ 0.0507$ &
        $[3,\, 17]$ &
        $[3,\, 29]$ &
        $[5,\, 31]$ &
        $[5,\, 31]$ \\
    \midrule
        $0.088$ &
        $0.0012 ~/~ 0.0363$ &
        $[4,\, 23]$ &
        $[4,\, 41]$ &
        $[7,\, 47]$ &
        $[7,\, 47]$ \\
    \bottomrule
    \end{tabular}
\end{table}

\begin{figure}
    \centering
    \includegraphics[width=0.75\textwidth]{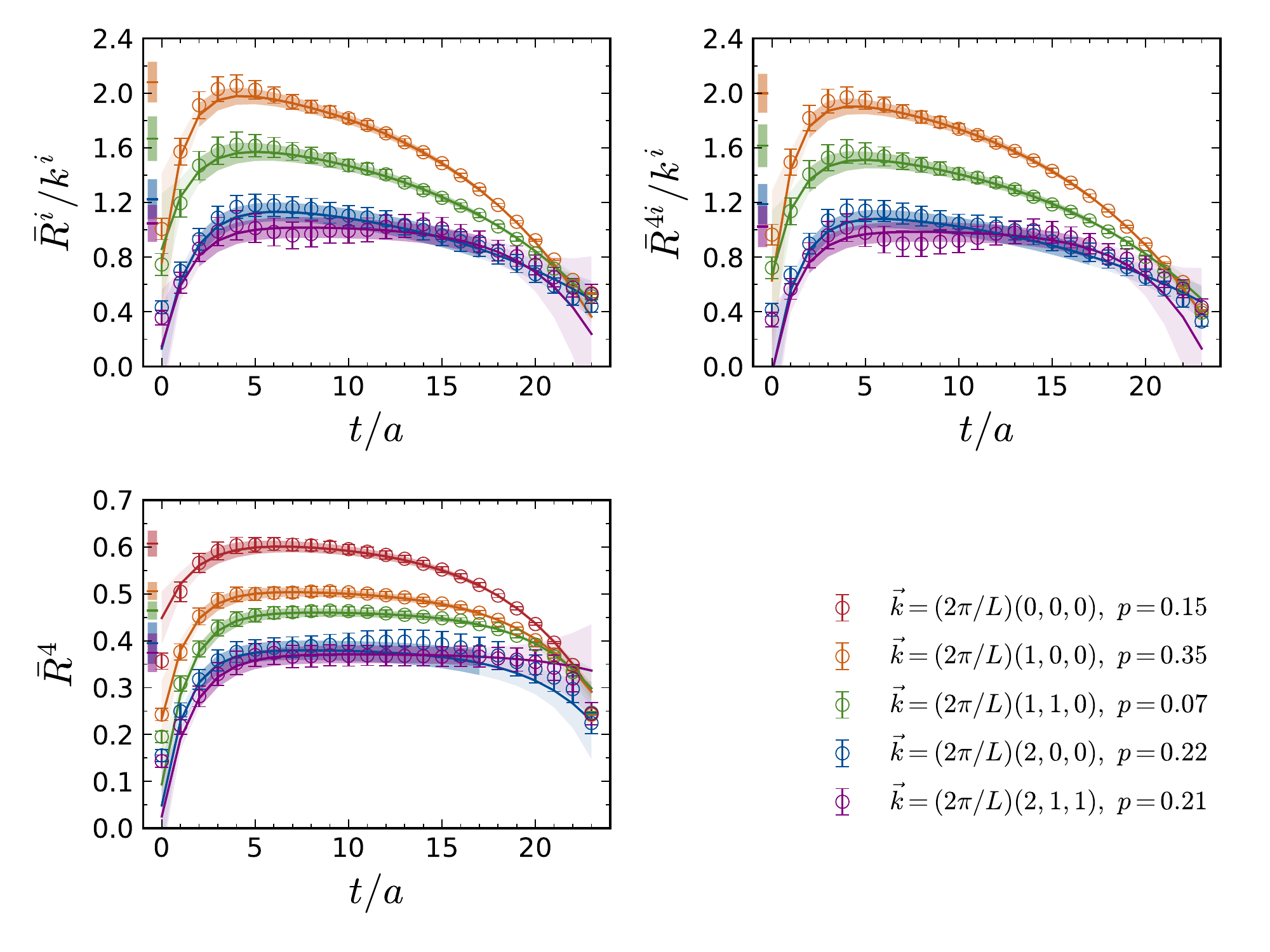}
    \caption{%
        Form factor ratios $\bar{R}$, fits, and unrenormalized results for
        $f_\perp$ (top left panel),
        $f_T$ (top right panel), and
        $f_\para$ (bottom left panel) in
        $B\to\pi$ at $a\simeq 0.088$~fm with $m_l/m_s=\text{physical}$.
        Fit bands are darker in the fit range, which is identical for all
        momenta and extends beyond that of the direct fits in both
        directions.
        Form-factor results (without $Z_J$) are indicated to the left of
        $t/a=0$; they are proportional to the green symbols in the leftmost
        column of Figure~\ref{fig:ffs}.
    }
    \label{fig:Rfit}
\end{figure}
\begin{figure}
    \centering
    \includegraphics[width=\textwidth]{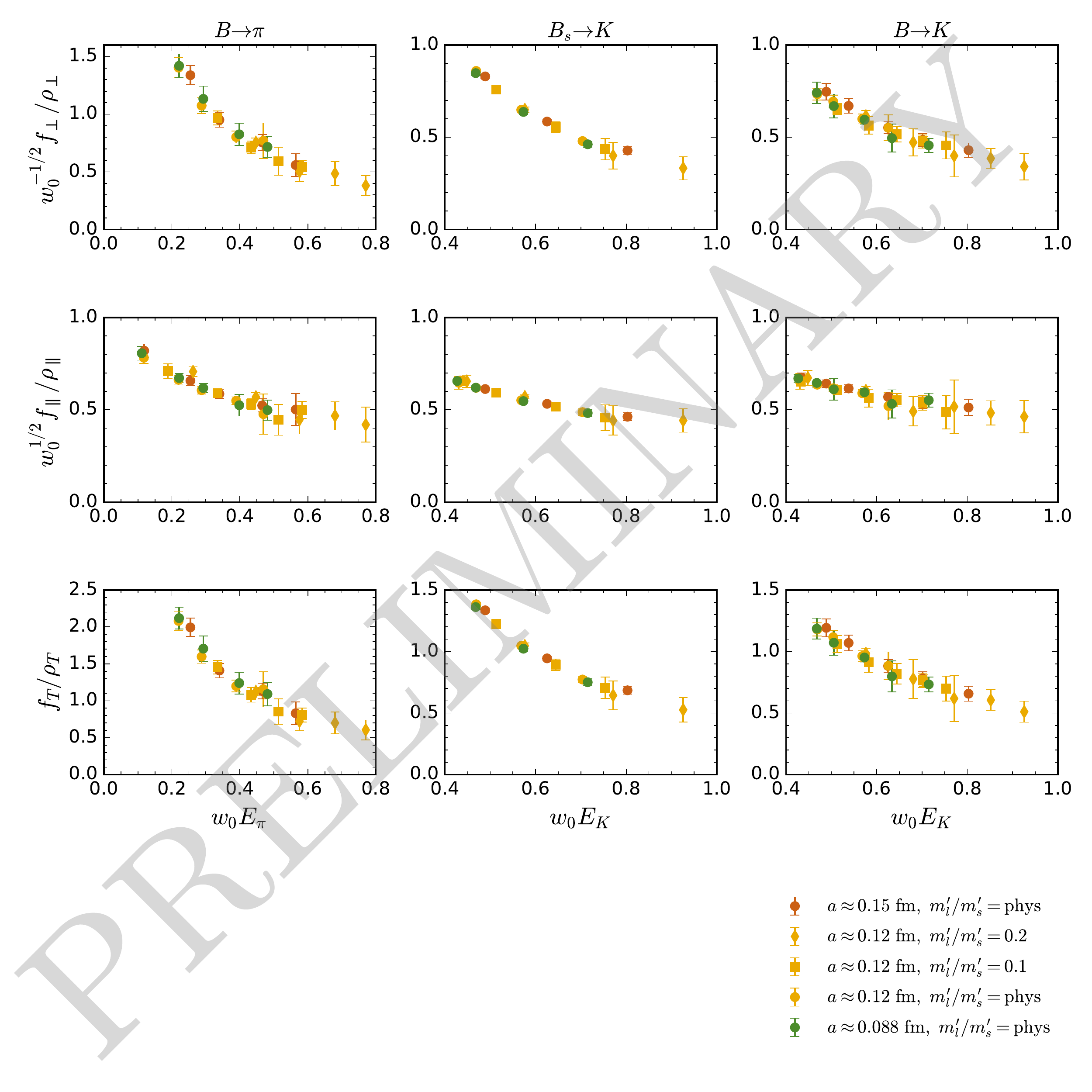}
    \caption{%
        Preliminary results for form factors
        $f_\perp$ (top),
        $f_\para$ (middle), and
        $f_T$ (bottom), in
        $B\to\pi$ (left),
        $B_s\to K$ (center), and
        $B\to K$ (right),
        without $\rho_J$.
    }
    \label{fig:ffs}
\end{figure}

Form factors may be extracted from fits to three-point correlation functions
with these forms:
\begin{align}
    C_3^{\mu(\nu)}(t, T;\bm{k}) &= \sum_{m,n=0}^{2N-1} (-1)^{m (t + 1)}
        (-1)^{n (T - t - 1)} \frac{Z_P^{(m)}}{E_P^{(m)}} A_{mn}^{\mu(\nu)}
        \frac{Z_B^{\dag (n)}}{M_B^{(n)}} e^{-E_P^{(m)} t}
        e^{-M_B^{(n)} (T - t)} ,
    \label{eq:C3fit}
\end{align}
where
$A_{mn}^{\mu(\nu)} = \left< P^{(m)}\midI J^{\mu(\nu)}\midI B^{(n)} \right>$.
Fitting directly to Eq.~(\ref{eq:C3fit}) yields the form factors as follows:
\begin{align}
    f_\perp(E_P) &= Z_{V^i} \frac{A_{00}^i(\bm{k})}{k^i} ,
    \label{eq:fperp} \\
    f_\para(E_P) &= Z_{V^4} A_{00}^4(\bm{k}) ,
    \label{eq:fpara} \\
    f_T(E_P)     &= Z_{T^{4i}} \frac{M_B + M_P}{\sqrt{2 M_B}}
                               \frac{A_{00}^{4 i}(\bm{k})}{k^i} ,
    \label{eq:fT}
\end{align}
where the current-renormalization factors $Z_J$ are defined in
Eq.~(\ref{eq:ZJ}). These fits are applied over ranges of $[t_\text{min}^P,
T - t_\text{min}^B]$ with loosely constrained fit priors. Alternatively, the
form factors may be obtained by computing ratios of three- to two-point
correlators, as introduced in Ref.~\cite{Bailey:2008wp}. For the form factors
presented in this report, we employ the latter procedure, as exemplified in
Figure~\ref{fig:Rfit}. The preliminary form factors for $B\to\pi$, $B_s\to K$,
and $B\to K$ are shown in Figure~\ref{fig:ffs}, where they include only the
nonperturbative factors in $Z_J$. The perturbative factors $\rho_J$ will be
blinded~\cite{Lattice:2015tia,Bailey:2015dka} once they are available. This will reduce subjective bias in our
analysis because the $\rho_J$ will be unblinded only after finalizing our
results.

Our analysis will proceed similarly to Refs.~\cite{Bailey:2008wp,Lattice:2015tia,Bailey:2015dka}.
We will perform a combined chiral-continuum extrapolation of the form factors.
Since the sea quarks are at physical mass in this work, chiral perturbation
theory will be used as an interpolation to correct for slight mistunings.
We do not expect $b$-quark mistuning to be significant because the parameters
are determined with greater precision in this work.
Next, we will construct a complete error budget, including uncertainties from
scale, dynamical quark masses, and current renormalization, as well as
effects from finite volume and the discretization of quarks and gluons.
Lastly, we will extrapolate to the full kinematic range accessible in
experiment by applying the model-independent $z$~expansion \cite{Boyd:1994tt,
Bourrely:2008za}.
It is then that the form factors will be unblinded and direct comparisons
with experiment can be made.

\section{Summary and outlook}\label{sec:outro}

We have presented preliminary results for the lattice-QCD calculation of the
form factors $f_\perp$, $f_\para$, and $f_T$ in the charged-current decays
$B_{(s)}\to\pi(K)\ell\nu_\ell$ and neutral-current decays
$B\to\pi(K)\ell^+\ell^-$. These calculations use HISQ ensembles from MILC,
which include physical quark masses.
We expect similar statistics and discretization errors to previous asqtad
analyses \cite{Bailey:2015dka, Lattice:2015tia} for ensembles at similar
lattice spacings. The main improvement in this work is the removal of the
errors due to the chiral extrapolation. Further improvement will be
necessary, both in the statistics for the nonperturbative renormalization
factors and in the availability of ensembles at finer lattice spacings.

Our analysis is meant to confront experiment in two ways. First, we will
determine $|V_{ub}|$ by combining with experimental results for the
charged-current decays, which will help to address the tension between
inclusive and exclusive determinations. Second, we will test for new physics
by comparing with experimental results for the neutral-current decays (e.g.,
branching fractions in $B\to K\mu^+\mu^-$), which will help to address the
tension between theory and experiment.

\subsection*{Acknowledgments}\label{sec:thanks}

This work was supported in part
by the U.S.~Department of Energy under grants
    No.~DE-AC05-06OR23177 (C.B.),
    No.~DE-SC0010120 (S.G.),
    No.~DE-SC0015655 (A.X.K., Z.G.),
    No.~DE-SC0010113 (Y.M.),
    No.~DE-FG02-13ER41976 (D.T.);
by the U.S.~National Science Foundation under grants
    PHY12-12389 (Y.L.),
    PHY14-14614 (C.D.), and
    PHY13-16748 and PHY16-20625 (R.S.);
by the Fermilab Distinguished Scholars Program (A.X.K.);
by German Excellence Initiative and the European Union Seventh Framework
    Program under grant agreement No.~291763 as well as the European Union's
    Marie Curie COFUND program (A.S.K.);
by the Blue Waters PAID program (Y.L.); and
by the Visiting Scholars Program of the Universities Research Association
    (Z.G., Y.L.).

Computations for this work were carried out with resources provided
by the USQCD Collaboration;
by the ALCF and NERSC, which are funded by the U.S.~Department of Energy; and
by NCAR, NCSA, NICS, TACC, and Blue Waters, which are funded through the
U.S.~National Science Foundation.
The Blue Waters sustained-petascale computing project is supported by the
National Science Foundation (awards OCI-0725070 and ACI-1238993) and the
state of Illinois. Blue Waters is a joint effort of the University of
Illinois at Urbana--Champaign and its National Center for Supercomputing
Applications.
Fermilab is operated by Fermi Research Alliance, LLC under Contract
No.~DE-AC02-07CH11359 with the United States Department of Energy, Office of
Science, Office of High Energy Physics. The United States Government retains
and the publisher, by accepting the article for publication, acknowledges
that the United States Government retains a non-exclusive, paid-up,
irrevocable, worldwide license to publish or reproduce the published form of
this manuscript, or allow others to do so, for United States Government
purposes.

\bibliography{lattice2017}

\end{document}